\documentclass[twocolumn,
prl
]{revtex4}

\usepackage{graphicx}
\usepackage{amssymb}
\usepackage{amsmath}
\usepackage{color}
\usepackage{comment}

\newcommand{\vect}[1]{{\mathbf #1}}

\def\lba{\left(}    \def\rba{\right)}
\def\lbc{\left[}    \def\rbc{\right]}

\newcommand{\bra}[1]{\langle\left.{#1}\right|}
\newcommand{\ket}[1]{\left|{#1}\right.\rangle}

\begin{document}


\title{Polaron-molecule transitions in a two-dimensional Fermi gas}

\author{Meera M. Parish}
\email{mmp24@cam.ac.uk} %
\affiliation{Cavendish Laboratory, JJ Thomson Avenue, Cambridge,
 CB3 0HE, UK} %

\date{\today}

\begin{abstract}
We address the problem of a single ``spin-down'' impurity atom interacting attractively with a spin-up Fermi gas in two dimensions (2D). We consider the case where the mass of the impurity is greater than or equal to the mass of a spin-up fermion.  Using a variational approach, we resolve the questions raised by previous studies and we show that there is, in fact, a transition between polaron and molecule (dimer) ground states in 2D. For the molecule state, we use a variational wave function 
with a single particle-hole excitation on the Fermi sea and we find that its energy matches that of the exact solution in the limit of infinite impurity mass. 
Thus, we expect the variational approach to provide a reliable tool for investigating 2D systems.
\end{abstract}


\maketitle

The impressive realization of Fermi systems in
ultracold atomic gases has greatly renewed interest in fundamental problems in 
pairing phenomena. For attractive interactions, the binding of fermions can alter the statistics of low-energy excitations and thus determine the low-energy behavior of Fermi systems such as Bose-Fermi mixtures and two-component Fermi gases. 
This phenomenon in Fermi systems is nicely captured by the  
fundamental problem: What is the ground state of a single impurity atom interacting attractively with a Fermi gas? This problem differs from the usual polaron problem of an impurity particle interacting with a background medium, since the medium is fermionic rather than bosonic. 
%
It also has connections with other Fermi systems: 
For a fermionic impurity, it corresponds to the extreme limit of spin imbalance in a two-component Fermi gas~\cite{partridge2006_2,shin2008,schirotzek2009,Nascimbene}. 

In three dimensions (3D), it is known that this ``spin-down''  impurity can undergo a sharp transition to a new ground state by binding a fermion from the spin-up Fermi sea~\cite{prokofiev2008,prokofiev2008_2}. For weak interactions, the impurity is initially dressed with density fluctuations of the Fermi gas, forming the so-called ``polaron'' state~\cite{chevy2006_2,combescot2008}.
Then, with increasing attraction, the impurity eventually binds a spin-up fermion to form a dimer or molecule state in the case of equal masses  ($m_\uparrow = m_\downarrow$)~\cite{mora2009,punk2009,combescot2009}. For a sufficiently light impurity ($m_\downarrow < m_\uparrow$), there is even the prospect of it binding \emph{two} spin-up fermions to form a dressed trimer~\cite{mathy2010}. 

However, the situation is less clear in lower dimensions, where quantum fluctuations are  
important. For 2D Fermi gases, Ref.~\cite{zollner2011} 
find no evidence of a polaron-molecule transition 
when particle-hole excitations of the Fermi sea are neglected in the molecule state. 
Instead, they find 
that the polaronic description persists into the regime of strong attraction, even when it is no longer accurate. 
This suggests that an extension of their variational \textit{ansatz} is required~\cite{zollner2011}.  
In this paper, we resolve this question and show that there is indeed a polaron-molecule transition in 2D, provided one treats the polaron and molecule on an equal footing and includes particle-hole excitations of the Fermi sea in \emph{both} variational wave functions. Here we consider a molecule wave function with a single particle-hole pair and we show that this is sufficient to reproduce the exact impurity energy in the limit of infinite impurity mass. 
We also show that the variational approach can never produce 
a polaron-molecule transition in 1D, even with the inclusion of particle-hole excitations, which is consistent with the finding that the polaronic description appears accurate 
across the whole range of interactions in 1D~\cite{1Dpolaron}. 
We expect our results to be important for the ongoing investigation of 2D atomic Fermi gases~\cite{2DFermi_expt,2DFermi_expt2}, as well as for other 2D Fermi systems in condensed matter such as electron-hole bilayers~\cite{parish2010}.

\paragraph{Model}
In the following, we consider a two-component
($\uparrow$, $\downarrow$) 2D system with short-range interspecies interactions, described by the Hamiltonian:
\begin{align}\label{eq:Ham}
 H = & \sum_{\vect{k}\sigma} \epsilon_{\vect{k}\sigma}
 c^\dag_{\vect{k}\sigma}c_{\vect{k}\sigma}
 + \frac{g}{\Omega}\sum_{\vect{k},\vect{k'},\vect{q}}
 c^\dag_{\vect{k}\uparrow}c^\dag_{\vect{k'}\downarrow}
 c_{\vect{k'}+\vect{q}\downarrow}c_{\vect{k}-\vect{q}\uparrow}~,
\end{align}
where $\epsilon_{\vect{k}\sigma} = \frac{\vect{k}^2}{2m_\sigma}$ (we
set $\hbar = 1$), $\Omega$ is the system area and $g$ is the attractive 
contact interaction. 
For a Fermi mixture, the Hamiltonian~\eqref{eq:Ham} completely describes the low-energy behavior, since 
Pauli exclusion suppresses $s$-wave interactions between the same species of fermion. For bosonic spin-down particles, we must also include intraspecies interactions between bosons. However, Eq.~\eqref{eq:Ham} is sufficient for this work since we restrict ourselves  to the problem of a single $\downarrow$-particle in a $\uparrow$ Fermi sea. 

To make our results independent of the UV cut-off for the short-range interaction,  
%
we use the energy eigenvalue equation for the two-body ($\uparrow$ and $\downarrow$) problem:  
%
\begin{align} \label{eq:2body}
-\frac{1}{g} & = \frac{1}{\Omega} \sum_{\vect{k}}^{\Lambda} 
\frac{1}{\varepsilon_B + \epsilon_{\vect{k}\uparrow} + \epsilon_{\vect{k}\downarrow}}~,
\end{align}
where $\Lambda$ is the momentum cut-off and $\varepsilon_B$ is the two-body binding energy, which always exists for an attractive interaction in 2D. This allows us to replace $g$ with $\varepsilon_B$ in our calculations and then take $\Lambda \to \infty$.

We parameterize the non-interacting Fermi gas with the Fermi energy $\varepsilon_F = k_F^2/2m_\uparrow$, where $k_F$ corresponds to the Fermi wave vector of the spin-up Fermi sea. Defining the mass ratio $r = m_\uparrow/m_\downarrow$, we focus on the regime $r\leq 1$; the limit $r\to 0$ can be compared with the exact solution~\cite{zollner2011},  
while the equal-mass case $r=1$ corresponds to current Fermi gas experiments~\cite{2DFermi_expt}.

\begin{figure}
\centering
\includegraphics[width=0.85\linewidth]{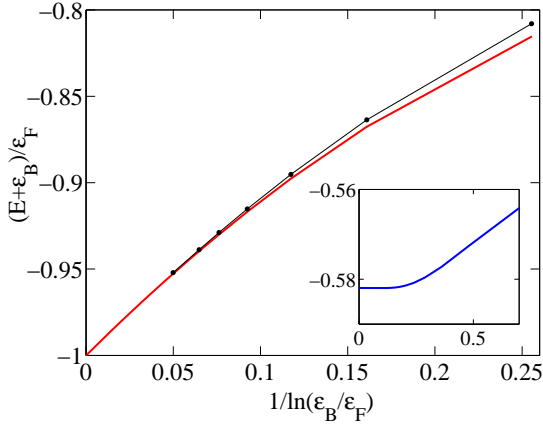}
\caption{(Color online) Energy of an impurity with infinite mass in the strong-coupling limit $\varepsilon_B /\varepsilon_F \gg 1 $. The thick (red) line corresponds to the exact solution \eqref{eq:exact}, while the thin (black) line with data points is the energy of the dressed molecule calculated using the variational wave function in Eq.~\eqref{eq:M4}. The energy calculated using the polaron variational wave function \eqref{eq:polaron} is depicted in the inset. In the limit $\varepsilon_B \to \infty$, the polaron energy approaches $E +\varepsilon_B = - \varepsilon_F/(e-1) \simeq -0.582\varepsilon_F$. Note that 
$E +\varepsilon_B$ depends linearly on $1/\varepsilon_B$ for the polaron in this limit, which gives rise to the apparent kink in the inset for the units $1/\ln(\varepsilon_B/\varepsilon_F)$.}
\label{fig:inf}
\end{figure}

\paragraph{Polaron}
For weak interactions (or, equivalently, large densities) where $\varepsilon_B/\varepsilon_F \ll 1$, 
the ground state is expected to be a polaron with the approximate wave function~\cite{chevy2006_2,combescot2008}:
\begin{align} \label{eq:polaron} 
\ket{P}  & =   \alpha_{0} c^\dag_{0\downarrow} \ket{N}
+  \frac{1}{\Omega} \sum_{\vect{k},\vect{q}} \alpha_{\vect{k}\vect{q}} c^\dag_{\vect{q} - \vect{k}\downarrow} 
c^\dag_{\vect{k}\uparrow} c_{\vect{q}\uparrow} \ket{N}
\end{align}
which contains just one particle-hole excitation on top of the Fermi sea $\ket{N}$ with $N$ spin-up fermions. We determine the ground state energy $E$ by minimizing $\bra{P} (H - E) \ket{P}$ with respect to the amplitudes $\alpha$, yielding the implicit equation:
\begin{align}\label{eq:polE}
E & =\sum_{|\vect{q}|<k_F} \lbc \frac{\Omega}{g} + \sum_{|\vect{k}|>k_F} \frac{1}{E_{\vect{k}\vect{q}}}  \rbc^{-1}
\end{align}
where $E_{\vect{k}\vect{q}} = -E + \epsilon_{\vect{q} - \vect{k}\downarrow} + \epsilon_{\vect{k}\uparrow} - \epsilon_{\vect{q}\uparrow}$. 
The real eigenstate will, of course, include terms with greater numbers of particle-hole pairs, 
but it has been argued that these terms approximately cancel in the variational equations for $E$~\cite{combescot2008}. Indeed, if we take $\alpha_{\vect{k}\vect{q}}$ to be independent of $\vect{q}$, i.e.\ $\alpha_{\vect{k}\vect{q}} \equiv \beta_{\vect{k}}$, then we obtain a closed set of equations for $\alpha_0$, $\beta_{\vect{k}}$ that do not involve higher-order terms. Thus, we can estimate the ``exactness'' of our variational approach by comparing the energy of the full variational wave function with that of our $\vect{q}$ approximation, where we simply replace $E_{\vect{k}\vect{q}}$ in Eq.~\eqref{eq:polE} with $\langle E_{\vect{k}\vect{q}} \rangle_{\vect{q}} = -E + \epsilon_{\vect{k}\downarrow} + \epsilon_{\vect{k}\uparrow}  + \varepsilon_F (r-1)/2$. Note that, for $r=1$, this is equivalent to replacing $E_{\vect{k}\vect{q}}$ with $E_{\vect{k}0}$, but it provides a better estimate of the energy for $r<1$. We find that the $\vect{q}$ approximation becomes better with decreasing $\varepsilon_B/\varepsilon_F$, implying that $\ket{P}$ is accurate at weak-coupling, as expected.

\paragraph{Dressed molecule}
For stronger interactions, we consider a molecule wave function with a single particle-hole excitation of the Fermi sea:
\begin{align}\notag
 \ket{M}  = &  \sum_{\vect{k}} \varphi_{\vect{k}}
c^\dag_{-\vect{k}\downarrow} c^\dag_{\vect{k}\uparrow} \ket{N-1}
 \\ \label{eq:M4} 
& 
+ \frac{1}{2\Omega}\sum_{\vect{k}\vect{k'}\vect{q}} \varphi_{\vect{k}\vect{k'}\vect{q}} c^\dag_{ \vect{q}-\vect{k} - \vect{k'}\downarrow} c^\dag_{\vect{k}\uparrow} c^\dag_{\vect{k'}\uparrow} c_{\vect{q}\uparrow}  \ket{N-1} 
\end{align}
Note that we use a Fermi sea of $N-1$ spin-up fermions in order to preserve particle number. 
By minimizing $\bra{M} (H - E) \ket{M}$ with respect to the $\varphi$'s, we obtain the simultaneous integral equations:
\begin{align} \label{eq:Meqn1}
\frac{\Omega}{g} & =  \sum_{\vect{k}}  \frac{1 + \sum_{\vect{q}} G(\vect{k},\vect{q}) }{E_M - \epsilon_{\vect{k}\uparrow} - \epsilon_{\vect{k}\downarrow}} \\
\notag
\frac{\Omega}{E_M - \epsilon_{\vect{k}\uparrow} - \epsilon_{\vect{k}\downarrow}}  &=  \ 
\lba \frac{\Omega}{g} +  \sum_{\vect{k'}} \frac{1}{E_{\vect{k}\vect{k'}\vect{q}}} \rba G(\vect{k},\vect{q})  \\
\label{eq:Meqn2}
& - \sum_{\vect{k'}} \frac{G(\vect{k'},\vect{q})}{E_{\vect{k}\vect{k'}\vect{q}}} - \frac{ \sum_{\vect{q'}} G(\vect{k},\vect{q'})}{E_M - \epsilon_{\vect{k}\uparrow} - \epsilon_{\vect{k}\downarrow}} 
\end{align}
%
where $E_M = E +\varepsilon_F$, $E_{\vect{k}\vect{k'}\vect{q}} = -E_M + \epsilon_{\vect{q}-\vect{k}-\vect{k'}\downarrow} + \epsilon_{\vect{k}\uparrow} + \epsilon_{\vect{k'}\uparrow} - \epsilon_{\vect{q}\uparrow}$ and
$G(\vect{k},\vect{q}) = \frac{1}{\Omega}\sum_{\vect{k'}} \varphi_{\vect{k}\vect{k'}\vect{q}}/\sum_{\vect{k'}}\varphi_\vect{k'}$. 
Here, it is understood that the hole momenta $\vect{q}$ satisfy $|\vect{q}| < k_F$ and particle momenta $\vect{k}$ satisfy $|\vect{k}| > k_F$. 
For $G(\vect{k},\vect{q}) = 0$, Eq.~\eqref{eq:Meqn1} recovers the equation for the ``bare'' molecule $|M_0^{(\vect{Q})}\rangle = \sum_{\vect{k}} \varphi_{\vect{k}}^{(\vect{Q})}
c^\dag_{\vect{Q}-\vect{k}\downarrow} c^\dag_{\vect{k}\uparrow} \ket{N-1}$ considered in \cite{zollner2011}, with center-of-mass momentum $\vect{Q}=0$. 
The bare state $\ket{M_0}$ 
never has a lower energy than $\ket{P}$ 
and is unable to reproduce the strong-coupling limit for 
$r=0$. In this case, the impurity energy is given exactly by~\cite{zollner2011}:
\begin{align}\label{eq:exact}
E & = -\varepsilon_B -\frac{2\varepsilon_F}{\pi} \int^1_{0} dk k \delta_0(k k_F)
\end{align}
where 
$\cot(\delta_0(k k_F)) = \ln(k^2\varepsilon_F/\varepsilon_B)/\pi$. 
For $\varepsilon_B \to \infty$, we have $\delta_0 \simeq \pi$ and thus $E \simeq -\varepsilon_B - \varepsilon_F$, which disagrees with the bare molecule energy of $- \varepsilon_B$. 
However, the inclusion of $G(\vect{k},\vect{q})$ has a dramatic effect on the molecule energy. To see this, we first assume $G(\vect{k},\vect{q})$ is small and drop the sums involving $G(\vect{k},\vect{q})$ in Eq.~\eqref{eq:Meqn2}. This allows us to solve for $G(\vect{k},\vect{q})$ and then substitute it into Eq.~\eqref{eq:Meqn1}. For $r\leq 1$, the solution of the resulting implicit equation yields $E +\varepsilon_F +\varepsilon_B \propto -\ln(\varepsilon_B)$, a correction which clearly diverges for $\varepsilon_B \to \infty$. This implies that $G(\vect{k},\vect{q})$ is not perturbative in $\varepsilon_B^{-1}$ and thus 
cannot be neglected in the strong-coupling regime. 

We have solved Eqs.~\eqref{eq:Meqn1}-\eqref{eq:Meqn2} numerically by discretizing momentum space and converting the problem into a matrix equation. 
The results for an impurity with infinite mass ($r=0$) are plotted in Fig.~\ref{fig:inf}. We see that the energy curve 
for $\ket{M}$ collapses onto the exact result as we take $\varepsilon_B \to \infty$, thus demonstrating that a molecule wave function with a single particle-hole pair is sufficient for capturing the strong-coupling limit. 
By contrast, the energy for the polaron $\ket{P}$  
rapidly approaches $E + \varepsilon_B =  - \varepsilon_F/(e-1)$ when $\varepsilon_B \to \infty$, as depicted in Fig.~\ref{fig:inf} (inset). This limit is obtained analytically by assuming $E + \varepsilon_B \ll \varepsilon_B$ and approximating the expression in the $\vect{q}$-sum in Eq.~\eqref{eq:polE} to get $E \simeq - \varepsilon_B \ln(1- 2\varepsilon_F/(E+\varepsilon_B))$. Taking exponentials and expanding in $(E + \varepsilon_B)/\varepsilon_B$ then gives the final result. 
For general $r \leq 1$, we can test the accuracy of $\ket{M}$ by comparing its energy with that of the $\vect{q}$ approximation, where we replace $E_{\vect{k}\vect{k'}\vect{q}}$ in Eq.~\eqref{eq:Meqn2} with $\langle  E_{\vect{k}\vect{k'}\vect{q}} \rangle_{\vect{q}} = -E_M + \epsilon_{\vect{k}+\vect{k'}\downarrow} + \epsilon_{\vect{k}\uparrow} + \epsilon_{\vect{k'}\uparrow} + \varepsilon_F (r-1)/2$.  As anticipated, we obtain excellent agreement for $\varepsilon_B/\varepsilon_F \gg 1$. 

\begin{figure}
\centering
\includegraphics[width=0.85\linewidth]{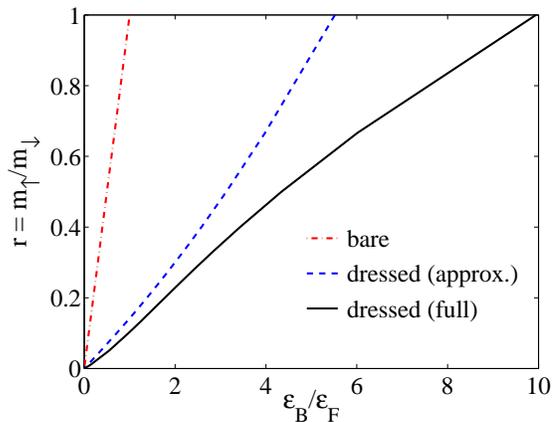}
\caption{(Color online) Polaron-molecule transitions that have been determined using three different variational approaches of increasing accuracy for the polaron and molecule. To the left (right) of each transition line, the polaron (molecule) is the ground state. The ``bare'' transition line was determined using the bare variational wave functions $\ket{P_0}$, $|M_0^{(\vect{Q}=0)}\rangle$  
and corresponds to a mean-field spinodal line (see text). 
The full calculation for the ``dressed'' variational wave functions $\ket{P}$, $\ket{M}$ yields the solid (black) line, while an approximate calculation that averages the dependence of the energy on the hole momentum $\vect{q}$ 
gives the dashed (blue) line.}
\label{fig:transition}
\end{figure}

\paragraph{Polaron-molecule transition}
The fact that the polaron state $\ket{P}$ has a much higher energy than the molecule state $\ket{M}$ in Fig.~\ref{fig:inf} already suggests that there must be a polaron-molecule transition at smaller $\varepsilon_B/\varepsilon_F$. Indeed, we even find a transition if we just consider the bare molecule $\ket{M_0}$ and the ``bare'' polaron state $\ket{P_0} = c^\dag_{0\downarrow} \ket{N}$, where particle-hole excitations of the Fermi sea are completely neglected.  
This ``bare'' binding transition, in fact, corresponds to a spinodal in the mean-field theory of a two-component Fermi gas~\cite{conduit2008} in the limit of full polarization. To see this, we note that the equation for the bare molecule $|M_0^{(\vect{Q})}\rangle$:
\begin{align}\label{eq:bare}
\frac{1}{g} & =  \frac{1}{\Omega} \sum_{\vect{k}}  \frac{1}{E_M - \epsilon_{\vect{k}\uparrow} - \epsilon_{\vect{Q}-\vect{k}\downarrow}}
\end{align}
is equivalent to the linearized mean-field gap equation for the superfluid order parameter at momentum $\vect{Q}$, with spin-up chemical potential $\mu_\uparrow = \varepsilon_F$ and spin-down chemical potential $\mu_\downarrow = E$. At the binding transition $E = 0$, the normal unpaired phase becomes linearly unstable to forming a paired Fermi superfluid. For $\vect{Q} = 0$, the transition occurs at $\varepsilon_{B}/\varepsilon_F = r$, as plotted in Fig.~\ref{fig:transition}. Note that, for an infinitely massive impurity ($r=0$), this implies that there is always a bound molecule when $\varepsilon_B > 0$, consistent with the exact solution~\eqref{eq:exact}. 

With the inclusion of particle-hole excitations, we find that the polaron-molecule transition survives provided we treat the polaron and molecule on an equal footing and consider wave functions $\ket{P}$ and $\ket{M}$. 
The effect of a single particle-hole pair is to shift the transition to higher $\varepsilon_B/\varepsilon_F$ for $r>0$, but we see in Fig.~\ref{fig:transition} that the behavior of the $\ket{P}$-$\ket{M}$ transition line is still qualitatively similar to that of the bare transition line.  
The $\vect{q}$ approximation gives a transition line that is closer to the full $\ket{P}$-$\ket{M}$ transition line than to the bare one, but it still underestimates $\varepsilon_B/\varepsilon_F$ for a given $r>0$ and it suggests that the variational approach is least quantitatively accurate for the transition at $r=1$. 
We find that the variational wave function $\ket{P}$ is less accurate than $\ket{M}$  at the transition 
and thus we expect the exact transition to lie at slightly higher $\varepsilon_B/\varepsilon_F$ for each $r>0$. 

Having established the existence of a polaron-molecule transition in 2D, it is natural to ask whether there is a similar scenario in 1D. The exact Bethe ansatz solution implies that there is no transition for $r=1$~\cite{McGuire1966}. Indeed, we can prove that there are no polaron-molecule transitions for general $r$ owing to the form of the singularities in the variational equations in 1D. 
Consider, first, the energy equation~\eqref{eq:bare} for $|M_0^{(\vect{Q})}\rangle$, with the sum over $\vect{k}$ replaced by an integral: $\frac{1}{\Omega}\sum_{\vect{k}} \to \int \frac{d^d k}{(2\pi)^d}$. 
For $Q \equiv |\vect{Q}| = k_F$, we see that the integrand has a singularity at the energy of the bare polaron $E = E_M - \varepsilon_F = 0$ when $\vect{k} = \vect{Q}$.
This singularity is integrable for dimensions $d=2,3$ and so the interaction $g$ is well-defined at this energy. However, for $d=1$, we have a divergent integral and thus the energy of $\ket{M_0}$ can never equal the energy of $\ket{P_0}$ for nonzero $g$ or $\varepsilon_B$, which means that the molecule never undergoes a transition to a polaron 
in this approximation. If we now consider the dressed polaron at finite momentum $\vect{p}$:
\begin{align} \notag
|P^{(\vect{p})} \rangle  & =   \alpha_{0}^{(\vect{p})} c^\dag_{\vect{p}\downarrow} \ket{N} 
+  \frac{1}{\Omega} \sum_{\vect{k},\vect{q}} \alpha_{\vect{k}\vect{q}}^{(\vect{p})} c^\dag_{\vect{p}+\vect{q} - \vect{k}\downarrow} 
c^\dag_{\vect{k}\uparrow} c_{\vect{q}\uparrow} \ket{N} 
\end{align}
then Eq.~\eqref{eq:polE} becomes
\begin{align}\notag
E - \epsilon_{\vect{p}\downarrow} & = \int \frac{d^d q}{(2\pi)^d}  \lbc \frac{1}{g} + \int \frac{d^d k}{(2\pi)^d} \frac{1}{E_{\vect{k}\vect{q}}^{(\vect{p})}}  \rbc^{-1}
\end{align}
where $E_{\vect{k}\vect{q}}^{(\vect{p})} = -E + \epsilon_{\vect{p} + \vect{q} - \vect{k}\downarrow} + \epsilon_{\vect{k}\uparrow} - \epsilon_{\vect{q}\downarrow}$. Here, there is a singularity in the $\vect{q}$ integral at the energy for $| M_0^{(\vect{p}-k_F\hat{p})} \rangle$ when $\vect{q} = -k_F \hat{p}$, which once again gives a divergent integral in 1D. Thus, we have $E \to -\infty$, i.e.\ $\varepsilon_B \to \infty$, as the energy of $\ket{P}$ approaches that of $\ket{M_0}$ from below. 
This singular structure is also present in higher-order variational wave functions in 1D, so that the higher-order molecule $\ket{M}$ always has a lower energy than $\ket{P}$ and so on. 
The absence of a 
transition in 1D is perhaps because there is no clear distinction between the statistics of a polaron and molecule in 1D, and so we can expect a crossover from the weak to strong-coupling regimes.
By contrast, in higher dimensions, a dilute gas of impurities will exhibit different behavior (Fermi liquid or Bose superfluid) depending on whether they form polarons or molecules. 


\paragraph{Discussion}
A remaining question is whether or not the dressed molecule can have a finite 
momentum $\vect{Q}$ in the ground state for $r>0$. It has been shown that this 
can give rise to a spatially-modulated superfluid for a low density of fermionic impurities~\cite{parish2010}. 
In the bare case, once $\varepsilon_B/\varepsilon_F < r (r+1)$, we find that the molecule $\ket{M_0}$ acquires a finite momentum 
%
$Q  = \frac{1+r}{r a} \sqrt{k_F a\sqrt{r}-1}$, 
where we have defined the lengthscale $a = 1/\sqrt{2m_r\varepsilon_B}$ using the reduced mass $m_r = (m_\downarrow^{-1} + m_\uparrow^{-1})^{-1}$. 
Thus, we find that the binding transition is actually at $\varepsilon_B/\varepsilon_F = r/(1+r)$, with 
$Q = k_F$ at this point, in contrast to the 3D case where we always obtain $Q=0$ in the regime $r \leq 1$. 
The physical interpretation in 2D is that the bare molecule binds when 
$\varepsilon_B$ equals the center-of-mass kinetic energy of a molecule at $Q=k_F$.  
However, further work is needed to determine whether or not this also occurs in the case of the dressed molecule and polaron. 

The molecule momentum $\vect{Q}$ at the polaron-molecule transition 
can affect the sharpness of the 
transition. For $Q=0$, 
we require the creation of an extra particle-hole pair at the Fermi surface when the molecule decays into a polaron (or vice versa) in order to conserve energy and momentum. Thus, we expect to have a ``first-order'' transition where we can have a metastable molecule (or polaron) beyond the transition~\cite{prokofiev2008,prokofiev2008_2,bruun2010}. However, for $Q=k_F$ at the transition, we do not require the creation of a particle-hole pair and thus the transition is continuous, with an excited molecule (polaron) decaying quickly into a polaron (molecule)  beyond the transition~\cite{mathy2010}. 

To connect with real 2D systems, we would ultimately like to know how the behavior of a single impurity extends to that of a finite density of spin-down particles. In particular, is the polaron-molecule transition thermodynamically stable? Clearly, this will depend on the statistics of the impurity. 
For a two-component Fermi gas, a mean-field analysis~\cite{conduit2008} of Eq.~\eqref{eq:Ham} shows that the bare polaron-molecule transition is preempted by a first-order phase transition from an unpaired normal phase to a paired superfluid at 
$\varepsilon_B/\varepsilon_F = \sqrt{1+r} - 1 $. 
However, mean-field theory is unlikely to give reliable results in 2D, so further work is required to better estimate the energy of the superfluid and/or determine the interactions between dressed impurities.  
In principle, atomic gas experiments can probe the polaron-molecule transition by measuring $|\alpha_0|^2$ for the polaron using RF spectroscopy~\cite{schirotzek2009} or by determining the effective mass of the impurity using low-lying compression modes of the gas~\cite{Nascimbene}.

To conclude, we have shown that the inclusion of a single particle-hole excitation in the variational wave function for the molecule is sufficient for reproducing the exact energy of an infinitely-massive impurity in the strong-coupling limit. Moreover, by using variational wave functions that capture both the weak and strong-coupling regimes, we have demonstrated that polaron-molecule transitions do exist in 2D but not in 1D.

\acknowledgments 
I am grateful to Francesca Marchetti, Charles Mathy and David Huse for useful discussions. 
This work was funded by the EPSRC under Grant No.\ EP/H00369X/1. 


\begin{thebibliography}{20}
\expandafter\ifx\csname natexlab\endcsname\relax\def\natexlab#1{#1}\fi
\expandafter\ifx\csname bibnamefont\endcsname\relax
  \def\bibnamefont#1{#1}\fi
\expandafter\ifx\csname bibfnamefont\endcsname\relax
  \def\bibfnamefont#1{#1}\fi
\expandafter\ifx\csname citenamefont\endcsname\relax
  \def\citenamefont#1{#1}\fi
\expandafter\ifx\csname url\endcsname\relax
  \def\url#1{\texttt{#1}}\fi
\expandafter\ifx\csname urlprefix\endcsname\relax\def\urlprefix{URL }\fi
\providecommand{\bibinfo}[2]{#2}
\providecommand{\eprint}[2][]{\url{#2}}

\bibitem[{\citenamefont{Partridge et~al.}(2006)\citenamefont{Partridge, Li,
  Liao, Hulet, Haque, and Stoof}}]{partridge2006_2}
\bibinfo{author}{\bibfnamefont{G.~B.} \bibnamefont{Partridge}},
  \bibnamefont{et~al.}, \bibinfo{journal}{Phys. Rev. Lett.}
  \textbf{\bibinfo{volume}{97}}, \bibinfo{pages}{190407}
  (\bibinfo{year}{2006}).

\bibitem[{\citenamefont{Shin et~al.}(2008)\citenamefont{Shin, Schunck,
  Schirotzek, and Ketterle}}]{shin2008}
\bibinfo{author}{\bibfnamefont{Y.}~\bibnamefont{Shin}},
  \bibinfo{author}{\bibfnamefont{C.~H.} \bibnamefont{Schunck}},
  \bibinfo{author}{\bibfnamefont{A.}~\bibnamefont{Schirotzek}},
  \bibnamefont{and} \bibinfo{author}{\bibfnamefont{W.}~\bibnamefont{Ketterle}},
  \bibinfo{journal}{Nature} \textbf{\bibinfo{volume}{451}},
  \bibinfo{pages}{689} (\bibinfo{year}{2008}).

\bibitem[{\citenamefont{Schirotzek et~al.}(2009)\citenamefont{Schirotzek, Wu,
  Sommer, and Zwierlein}}]{schirotzek2009}
\bibinfo{author}{\bibfnamefont{A.}~\bibnamefont{Schirotzek}},
  \bibinfo{author}{\bibfnamefont{C.-H.} \bibnamefont{Wu}},
  \bibinfo{author}{\bibfnamefont{A.}~\bibnamefont{Sommer}}, \bibnamefont{and}
  \bibinfo{author}{\bibfnamefont{M.~W.} \bibnamefont{Zwierlein}},
  \bibinfo{journal}{Phys. Rev. Lett.} \textbf{\bibinfo{volume}{102}},
  \bibinfo{eid}{230402} (\bibinfo{year}{2009}).

\bibitem[{\citenamefont{Nascimb\`ene et~al.}(2009)\citenamefont{Nascimb\`ene,
  Navon, Jiang, Tarruell, Teichmann, McKeever, Chevy, and
  Salomon}}]{Nascimbene}
\bibinfo{author}{\bibfnamefont{S.}~\bibnamefont{Nascimb\`ene}},
  \bibnamefont{et~al.}, \bibinfo{journal}{Phys. Rev. Lett.}
  \textbf{\bibinfo{volume}{103}}, \bibinfo{pages}{170402}
  (\bibinfo{year}{2009}).

\bibitem[{\citenamefont{Prokof'ev and
  Svistunov}(2008{\natexlab{a}})}]{prokofiev2008}
\bibinfo{author}{\bibfnamefont{N.}~\bibnamefont{Prokof'ev}} \bibnamefont{and}
  \bibinfo{author}{\bibfnamefont{B.}~\bibnamefont{Svistunov}},
  \bibinfo{journal}{Phys. Rev. B} \textbf{\bibinfo{volume}{77}},
  \bibinfo{eid}{020408} (\bibinfo{year}{2008}{\natexlab{a}}).

\bibitem[{\citenamefont{Prokof'ev and
  Svistunov}(2008{\natexlab{b}})}]{prokofiev2008_2}
\bibinfo{author}{\bibfnamefont{N.~V.} \bibnamefont{Prokof'ev}}
  \bibnamefont{and} \bibinfo{author}{\bibfnamefont{B.~V.}
  \bibnamefont{Svistunov}}, \bibinfo{journal}{Phys. Rev. B}
  \textbf{\bibinfo{volume}{77}}, \bibinfo{eid}{125101}
  (\bibinfo{year}{2008}{\natexlab{b}}).

\bibitem[{\citenamefont{Chevy}(2006)}]{chevy2006_2}
\bibinfo{author}{\bibfnamefont{F.}~\bibnamefont{Chevy}},
  \bibinfo{journal}{Phys. Rev. A} \textbf{\bibinfo{volume}{74}},
  \bibinfo{pages}{063628} (\bibinfo{year}{2006}).

\bibitem[{\citenamefont{Combescot and Giraud}(2008)}]{combescot2008}
\bibinfo{author}{\bibfnamefont{R.}~\bibnamefont{Combescot}} \bibnamefont{and}
  \bibinfo{author}{\bibfnamefont{S.}~\bibnamefont{Giraud}},
  \bibinfo{journal}{Phys. Rev. Lett.} \textbf{\bibinfo{volume}{101}},
  \bibinfo{eid}{050404} (\bibinfo{year}{2008}).

\bibitem[{\citenamefont{Mora and Chevy}(2009)}]{mora2009}
\bibinfo{author}{\bibfnamefont{C.}~\bibnamefont{Mora}} \bibnamefont{and}
  \bibinfo{author}{\bibfnamefont{F.}~\bibnamefont{Chevy}},
  \bibinfo{journal}{Phys. Rev. A} \textbf{\bibinfo{volume}{80}},
  \bibinfo{eid}{033607} (\bibinfo{year}{2009}).

\bibitem[{\citenamefont{Punk et~al.}(2009)\citenamefont{Punk, Dumitrescu, and
  Zwerger}}]{punk2009}
\bibinfo{author}{\bibfnamefont{M.}~\bibnamefont{Punk}},
  \bibinfo{author}{\bibfnamefont{P.~T.} \bibnamefont{Dumitrescu}},
  \bibnamefont{and} \bibinfo{author}{\bibfnamefont{W.}~\bibnamefont{Zwerger}},
  \bibinfo{journal}{Phys. Rev. A} \textbf{\bibinfo{volume}{80}},
  \bibinfo{eid}{053605} (\bibinfo{year}{2009}).

\bibitem[{\citenamefont{Combescot et~al.}(2009)\citenamefont{Combescot, Giraud,
  and Leyronas}}]{combescot2009}
\bibinfo{author}{\bibfnamefont{R.}~\bibnamefont{Combescot}},
  \bibinfo{author}{\bibfnamefont{S.}~\bibnamefont{Giraud}}, \bibnamefont{and}
  \bibinfo{author}{\bibfnamefont{X.}~\bibnamefont{Leyronas}},
  \bibinfo{journal}{Europhys. Lett.} \textbf{\bibinfo{volume}{88}},
  \bibinfo{pages}{60007} (\bibinfo{year}{2009}).

\bibitem[{\citenamefont{Mathy et~al.}()\citenamefont{Mathy, Parish, and
  Huse}}]{mathy2010}
\bibinfo{author}{\bibfnamefont{C.~J.~M.} \bibnamefont{Mathy}},
  \bibinfo{author}{\bibfnamefont{M.~M.} \bibnamefont{Parish}},
  \bibnamefont{and} \bibinfo{author}{\bibfnamefont{D.~A.} \bibnamefont{Huse}},
  \bibinfo{note}{arXiv:1002.0101}.

\bibitem[{\citenamefont{Z\"ollner et~al.}(2011)\citenamefont{Z\"ollner, Bruun,
  and Pethick}}]{zollner2011}
\bibinfo{author}{\bibfnamefont{S.}~\bibnamefont{Z\"ollner}},
  \bibinfo{author}{\bibfnamefont{G.~M.} \bibnamefont{Bruun}}, \bibnamefont{and}
  \bibinfo{author}{\bibfnamefont{C.~J.} \bibnamefont{Pethick}},
  \bibinfo{journal}{Phys. Rev. A} \textbf{\bibinfo{volume}{83}},
  \bibinfo{pages}{021603} (\bibinfo{year}{2011}).

\bibitem[{\citenamefont{Giraud and Combescot}(2009)}]{1Dpolaron}
\bibinfo{author}{\bibfnamefont{S.}~\bibnamefont{Giraud}} \bibnamefont{and}
  \bibinfo{author}{\bibfnamefont{R.}~\bibnamefont{Combescot}},
  \bibinfo{journal}{Phys. Rev. A} \textbf{\bibinfo{volume}{79}},
  \bibinfo{pages}{043615} (\bibinfo{year}{2009}).

\bibitem[{\citenamefont{Martiyanov et~al.}(2010)\citenamefont{Martiyanov,
  Makhalov, and Turlapov}}]{2DFermi_expt}
\bibinfo{author}{\bibfnamefont{K.}~\bibnamefont{Martiyanov}},
  \bibinfo{author}{\bibfnamefont{V.}~\bibnamefont{Makhalov}}, \bibnamefont{and}
  \bibinfo{author}{\bibfnamefont{A.}~\bibnamefont{Turlapov}},
  \bibinfo{journal}{Phys. Rev. Lett.} \textbf{\bibinfo{volume}{105}},
  \bibinfo{pages}{030404} (\bibinfo{year}{2010}).

\bibitem[{\citenamefont{Fr\"ohlich et~al.}(2011)\citenamefont{Fr\"ohlich, Feld,
  Vogt, Koschorreck, Zwerger, and K\"ohl}}]{2DFermi_expt2}
\bibinfo{author}{\bibfnamefont{B.}~\bibnamefont{Fr\"ohlich}},
  \bibnamefont{et~al.}, \bibinfo{journal}{Phys. Rev. Lett.}
  \textbf{\bibinfo{volume}{106}}, \bibinfo{pages}{105301}
  (\bibinfo{year}{2011}).

\bibitem[{\citenamefont{Parish et~al.}()\citenamefont{Parish, Marchetti, and
  Littlewood}}]{parish2010}
\bibinfo{author}{\bibfnamefont{M.~M.} \bibnamefont{Parish}},
  \bibinfo{author}{\bibfnamefont{F.~M.} \bibnamefont{Marchetti}},
  \bibnamefont{and} \bibinfo{author}{\bibfnamefont{P.~B.}
  \bibnamefont{Littlewood}}, \bibinfo{note}{arXiv:1009.1420}.

\bibitem[{\citenamefont{Conduit et~al.}(2008)\citenamefont{Conduit, Conlon, and
  Simons}}]{conduit2008}
\bibinfo{author}{\bibfnamefont{G.~J.} \bibnamefont{Conduit}},
  \bibinfo{author}{\bibfnamefont{P.~H.} \bibnamefont{Conlon}},
  \bibnamefont{and} \bibinfo{author}{\bibfnamefont{B.~D.}
  \bibnamefont{Simons}}, \bibinfo{journal}{Phys. Rev. A}
  \textbf{\bibinfo{volume}{77}}, \bibinfo{pages}{053617}
  (\bibinfo{year}{2008}).

\bibitem[{\citenamefont{McGuire}(1966)}]{McGuire1966}
\bibinfo{author}{\bibfnamefont{J.~B.} \bibnamefont{McGuire}},
  \bibinfo{journal}{J. Math. Phys.} \textbf{\bibinfo{volume}{7}},
  \bibinfo{pages}{123} (\bibinfo{year}{1966}).

\bibitem[{\citenamefont{Bruun and Massignan}(2010)}]{bruun2010}
\bibinfo{author}{\bibfnamefont{G.~M.} \bibnamefont{Bruun}} \bibnamefont{and}
  \bibinfo{author}{\bibfnamefont{P.}~\bibnamefont{Massignan}},
  \bibinfo{journal}{Phys. Rev. Lett.} \textbf{\bibinfo{volume}{105}},
  \bibinfo{pages}{020403} (\bibinfo{year}{2010}).

\end{thebibliography}

\end{document}